\documentclass[a4paper]{article}
\usepackage{amssymb}
\usepackage{amsfonts}
\usepackage{amsmath}
\usepackage{amsthm}
\usepackage{amsopn}
\usepackage{bbm}
\usepackage{multicol}
\usepackage{a4wide}

\newcommand{\slnc}{\mathfrak{sl}(N+1,\mathbb{C})}
\newcommand{\C}{\mathbb{C}}

\newcommand{\np}{\mathfrak{n}_+}
\newcommand{\nm}{\mathfrak{n}_-}
\newcommand{\g}{\mathfrak{g}}
\newcommand{\h}{\mathfrak{h}}
\newcommand{\ra}{\Phi^\ap}
\newcommand{\rb}{\Phi^\bp}
\newcommand{\ap}{\mathfrak{a}}
\newcommand{\am}{\widetilde{\mathfrak{a}}}
\newcommand{\bp}{\mathfrak{b}}
\newcommand{\bm}{\widetilde{\mathfrak{b}}}
\newcommand{\ca}{\mathfrak{c}_1}
\newcommand{\cb}{\mathfrak{c}_2}
\newcommand{\ordb}{\Xi}
\newcommand{\posr}{\Phi_+}
\newcommand{\negr}{\Phi_-}
\newcommand{\rset}{\Phi}
\newcommand{\baser}{\Delta}
\newcommand{\up}{U}
\newcommand{\um}{\widetilde{U}}
\newcommand{\vp}{V}
\newcommand{\vm}{\widetilde{V}}

\newcommand{\ad}[1]{\mathrm{ad}_{#1}}

\DeclareMathOperator{\spn}{span}
\DeclareMathOperator{\htr}{ht}

\newtheorem{Theorem}{Theorem}[section]

\newtheorem{Lemma}[Theorem]{Lemma}
\newtheorem{Corollary}[Theorem]{Corollary}
\newtheorem{Definition}[Theorem]{Definition}

\begin{document}
\title{Wei-Norman equations for classical groups}

\author{
  Szymon Charzy\'nski\\
  Faculty of Mathematics and Natural Sciences,\\ 
  Cardinal Stefan Wyszy\'nski University,\\
  ul. W\'oycickiego 1/3, 01-938 Warszawa, Poland\\
  \texttt{szycha@cft.edu.pl}\\
  \\
  Marek Ku\'{s}\\
  Center for Theoretical Physics, Polish Academy of Sciences,\\
  Aleja Lotnik\'ow 32/46, 02-668 Warszawa, Poland\\
  \texttt{marek.kus@cft.edu.pl}
       }
\date{December 16, 2013}

\maketitle

\begin{abstract}
We show that the non-linear autonomus Wei-Norman equations, expressing the
solution of a linear system of non-autonomous equations on a Lie algebra, can
be reduced to the hierarchy of matrix Riccati equations in the case of all
classical simple Lie algebras. The result generalizes our previous one
concerning the complex Lie algebra of the special linear group. We show that it cannot be extended to all simple Lie algebras, in particular to the exceptional $G_2$ algebra.

\end{abstract}


\section{Introduction}
For a non-autonomous system of $N$ linear differential equations
\begin{equation}\label{genlin}
\frac{d}{dt}\mathbf{x}(t)=M(t)\mathbf{x}(t),
\end{equation}
where $\mathbf{x}(t)$ is an $N$ dimensional vector and the $N\times N$
coefficient matrix is time-dependent it is, in general, difficult to find a
solution in a finite form (not invoking a series expansion). A possible
non-commutativity of the coefficient matrices $M(t)$ and $M(t^\prime)$
calculated in different times $t$, $t^\prime$ prevents, namely, an explicit
calculation of the ordered time product of exponentials of the time dependent
matrix variable.

Wei and Norman \cite{wei63,wei64} proposed a method for solving such a system
of equations by transforming it to an autonomous, albeit nonlinear system using
Lie-algebras techniques. Roughly speaking, the method consists of writing
$M(t)$ as a linear combination with time dependent coefficients of generators
$X_i$ of a Lie algebra, $M(t)=\sum_ia_i(t)X_i$, and looking for the solution in
terms of a product of exponentials of the generators (i.e.\ elements of the
corresponding Lie group), $\mathbf{x}(t)=\prod_i\exp(u_i(t)X_i)\mathbf{x}(0)$.
The resulting system of nonlinear equations derived from the original,
non-autonomous, linear one (\ref{genlin}) connects the unknown functions
$u_i(t)$ with the coefficients $a_k(t)$.

In \cite{charzynski13} we have shown that using the Wei-Norman technique
\cite{wei63,wei64} in the unitary case, i.e.\ when the solution of the linear
system is given by a unitary evolution operator, the nonlinear system by an
appropriate choice of ordering can be reduced to a hierarchy of matrix Riccati
equations. To this end we have considered a general linear non-autonomous
dynamical system on the special linear group $SL(N+1,\C)$. The algebra $\slnc$
is one of the classical Lie algebras, of the type denoted by $A_N$. In the
present paper we generalize the method developed in \cite{charzynski13} to all
classical Lie algebras, namely the algebras of type $A_N$, $B_N$, $C_N$ and
$D_N$. In particular we show that in all classical cases the resulting
nonlinear system can be always reduced to a hierarchy of matrix Riccati
equations if we chose an appropriate ordered basis of the underlying
Lie-algebra. Thus, we expand the applicability of the Wei-Norman method from
the unitary and special linear groups to orthogonal and symplectic groups.

To exhibit the Riccati structure of the Wei-Norman equations in the case of
$A_N$ algebras considered in \cite{charzynski13} we proved several facts
concerning their structural properties, among them two crucial ones. The first
establishing a decomposition of the $\slnc$ Lie algebra into a semidirect sum
of Abelian subalgebras and the second concerning the nilpotency of the adjoint
endomorphism. Here we show that these two observations remain valid for all
classical simple Lie algebras. Interestingly they cannot be extended to all
simple Lie algebras - we show this in the case of an exceptional Lie algebra
$G_2$.

\section{The Wei-Norman method}
\label{sec:WN}

We briefly recall our analysis of the Wei-Norman method presented in greater
detail in \cite{charzynski13}. Let $G$ be an $n$-dimensional Lie group and
$\mathfrak{g}$ - its Lie algebra. We assume in the following that
$\mathfrak{g}$ is a simple complex classical Lie algebra. Let also
$\mathbb{R}\ni t\mapsto M(t)\in\mathfrak{g}$ be a curve in $\mathfrak{g}$ and
$K(t)$ - a curve in $G$ given by the differential equation
\begin{equation}\label{precession}
\frac{d}{dt}K(t)=M(t)K(t),\quad K(0)=I.
\end{equation}
In $\mathfrak{g}$ we choose some basis $X_k$, $k=1,\ldots,n$ in which $M(t)$
takes the form
\begin{equation}\label{Mdec}
M(t)=\sum_{k=1}^na_k(t)X_k.
\end{equation}
We look for the solution $K(t)$ in the form
\begin{equation}\label{coord2}
K(t)=\prod_{k=1}^n\exp\big(u_k(t)X_k\big),
\end{equation}
involving $n$ unknown functions $u_k$ and this is the original idea of Wei and
Norman \cite{wei63}, to use product of exponentials instead of most commonly
used exponential of a linear combination of Lie algebra generators.

Substituting (\ref{Mdec}) and (\ref{coord2}) into (\ref{precession}) we
straightforwardly arrive at the following system of coupled differential
equations for $u_k$ in terms of $a_l$ (for a detailed derivation see
\cite{charzynski13}):
\begin{equation}\label{eq-wn0}
\sum_{k=1}^na_k X_k=\sum_{l=1}^nu_l^\prime\prod_{k<l}
\exp(u_k\mathrm{ad}_{X_k})\cdot X_l,
\end{equation}
where $\mathrm{ad}_X=[X,.]$ is the adjoint action of $\mathfrak{g}$ on itself.

The equation (\ref{eq-wn0}) can be rewritten as
\begin{equation}\label{aA}
\sum_{j=1}^n a_jX_j
=\sum_{j=1}^n\left(\sum_{l=1}^n A^{(l)}_{jl} u_l^\prime\right)X_j,
\end{equation}
where we denoted
\begin{equation}\label{Aldef}
A^{(l)}=\prod_{k<l}\exp( u_k\mathrm{ad}_{X_k}).
\end{equation}
Comparing corresponding coefficients of both sides of (\ref{aA}) we obtain
\begin{displaymath}
a_j=\sum_{l=1}^n A^{(l)}_{jl} u_l^\prime,
\end{displaymath}
or in a compact form,
\begin{equation}\label{eq-wn2a}
\mathbf{a}=A\mathbf{u}^\prime,
\end{equation}
where $A$ is an $n\times n$ matrix with elements $A_{jl}=A^{l}_{jl}$, i.e., its
$l$-th column is equal to the $l$-th column of the matrix $A^{(l)}$, c.f.
(\ref{Aldef}), and $\mathbf{a}$ and $\mathbf{u}$ are the vectors of the
coefficients of $M$ in (\ref{Mdec}) and the unknowns $u_k$. It was shown in
\cite{charzynski13} that $A$ is invertible at least locally, so we obtain a
system of (nonlinear) differential equations solved for the first derivatives
\begin{equation}\label{eq-wn3}
\mathbf{u}^\prime=A^{-1}\mathbf{a}.
\end{equation}

\section{General properties of simple complex Lie algebras}
\label{sec:slN}

In this section we give a brief summary of commonly known properties of Lie
algebras crucial for our purposes (see for example \cite{humphreys}) and fix
the notation.

Each complex simple Lie algebra $\mathfrak{g}$ can be decomposed into the
root spaces with respect to a chosen Cartan subalgebra $\mathfrak{h}$ (a
maximal commutative subalgebra of $\mathfrak{g}$),
\begin{equation}\label{rootdecomp}
\mathfrak{g}=\mathfrak{h}\oplus\mathop{\bigoplus}\limits_{\alpha\in\rset}
\mathfrak{g}_\alpha,
\end{equation}
where the one-dimensional \textit{root spaces} are defined as
\begin{displaymath}
\mathfrak{g}_\alpha:=\{X\in\mathfrak{g}:[H,X]=\alpha(H)X\;
\forall H\in\mathfrak{h}\}.
\end{displaymath}
The linear forms $\alpha\in\h^*$ (the dual space to the algebra $\mathfrak{h}$)
are called \emph{roots} and the element $X_\alpha$ spanning the subspace
$\mathfrak{g}_\alpha$ is called the \emph{root vector}. The number
$N=\dim\mathfrak{h}$ is called the rank of $\mathfrak{g}$. The set of roots we
will denote by $\rset$. The elements of $\rset$ span $\h^*$, which is an
$N$-dimensional Euclidean space with the scalar product induced by the Killing
form on $\h$. Among the roots we can find a set $\baser$ of $N$ linearly
independent \textit{positive simple roots},
$\baser=\{\alpha_1,\ldots\alpha_N\}$, such that for each $\alpha\in\rset$,
\begin{displaymath}
\alpha=\sum_{i=1}^N n_i\alpha_i,
\end{displaymath}
where either all $n_i$ are nonnegative (such roots $\alpha$ form a set of
\textit{positive roots} $\posr$) or all $n_i$ are non-positive (such roots
$\alpha$ form the set $\negr$ of negative roots). There is a one-to-one
correspondence between positive and negative roots: for each positive root
$\alpha$ the is a negative one $-\alpha$.

The root spaces have the following property
\begin{equation}\label{commg}
[\mathfrak{g}_\alpha,\mathfrak{g}_\beta]\subset\mathfrak{g}_{\alpha+\beta},
\end{equation}
(for $\beta=-\alpha$ we have
$[\mathfrak{g}_\alpha,\mathfrak{g}_{-\alpha}]\subset\mathfrak{h}$).

In terms of positive and negative roots the decomposition (\ref{rootdecomp})
can be rewritten as
\begin{equation}\label{cartan}
\g=\nm\oplus\h\oplus\np, \quad
\mathfrak{n}_\pm=\mathop{\bigoplus}\limits_{\alpha\in\rset_\pm}
\mathfrak{g}_\alpha.
\end{equation}

\section{Decomposition of  a classical Lie algebra of rank $N$ into a sum of $2N+1$ commuting subalgebras}

As mentioned in the Introduction the possibility of writing the Wei-Norman
equations in the form of a hierarchy of matrix Riccati equations hinges
crucially on two observations concerning the structure of the underlying Lie
algebra and its adjoint endomorphism. In the present section we state and prove
the first of them for an arbitrary classical simple Lie algebra - we show how
to decompose such an algebra into a sum of commutative subalgebras.

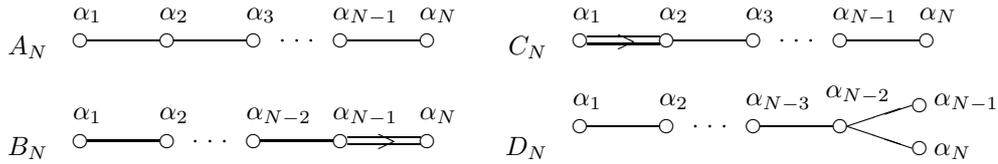
\begin{figure}[h]
\newcommand{\circrad}{1}
\newcommand{\circdiam}{2}
\newcommand{\linelength}{10}
\newcommand{\veclength}{12}
\newcommand{\vs}{5}
\setlength{\unitlength}{.95mm}
\renewcommand{\arraystretch}{3}
\begin{tabular}{rlcrl}
$A_N$ & \begin{picture}(50, 10)
  \put(\circrad, 2){\line(1, 0){\linelength}}
  \put(13, 2){\line(1, 0){\linelength}}
  \put(37, 2){\line(1, 0){\linelength}}
  \multiput(28,2)(2,0){3}{\circle*{.5}}
  \multiput(0, 2)(\veclength, 0){5}{\circle{\circdiam}}
  \put(-1,\vs){$\alpha_1$}
  \put(11,\vs){$\alpha_2$}
  \put(23,\vs){$\alpha_3$}
  \put(35,\vs){$\alpha_{N-1}$}
  \put(47,\vs){$\alpha_N$}
\end{picture} &\qquad&
$C_N$ &  \begin{picture}(50, 10)
  \put(\circrad, 2.5){\line(1, 0){\linelength}}
  \put(\circrad, 1.5){\line(1, 0){\linelength}}
  \put(5, 0.95){$>$}
  \put(13, 2){\line(1, 0){\linelength}}
  \put(37, 2){\line(1, 0){\linelength}}
  \multiput(28,2)(2,0){3}{\circle*{.5}}
  \multiput(0, 2)(\veclength, 0){5}{\circle{\circdiam}}
  \put(-1,\vs){$\alpha_1$}
  \put(11,\vs){$\alpha_2$}
  \put(23,\vs){$\alpha_3$}
  \put(35,\vs){$\alpha_{N-1}$}
  \put(47,\vs){$\alpha_N$}
\end{picture}\\
$B_N$ &  \begin{picture}(50, 10)
  \put(37, 2.5){\line(1, 0){\linelength}}
  \put(37, 1.5){\line(1, 0){\linelength}}
  \put(41, 0.95){$>$}
  \put(\circrad, 2){\line(1, 0){\linelength}}
  \put(25, 2){\line(1, 0){\linelength}}
  \multiput(16,2)(2,0){3}{\circle*{.5}}
  \multiput(0, 2)(\veclength, 0){5}{\circle{\circdiam}}
  \put(-1,\vs){$\alpha_1$}
  \put(11,\vs){$\alpha_2$}
  \put(23,\vs){$\alpha_{N-2}$}
  \put(35,\vs){$\alpha_{N-1}$}
  \put(47,\vs){$\alpha_N$}
\end{picture} & &
$D_N$ &  \begin{picture}(55, 10)
  \put(37, 4){\line(3, 1){9}}
  \put(37, 4){\line(3, -1){9}}
  \put(47, 7){\circle{\circdiam}}
  \put(47, 1){\circle{\circdiam}}
  \put(\circrad, 4){\line(1, 0){\linelength}}
  \put(25, 4){\line(1, 0){\linelength}}
  \multiput(16,4)(2,0){3}{\circle*{.5}}
  \multiput(0, 4)(\veclength, 0){4}{\circle{\circdiam}}
  \put(-1,7){$\alpha_1$}
  \put(11,7){$\alpha_2$}
  \put(23,7){$\alpha_{N-3}$}
  \put(34,8){$\alpha_{N-2}$}
  \put(49,7){$\alpha_{N-1}$}
  \put(49,0){$\alpha_{N}$}
\end{picture}
\end{tabular}
\caption{Dynkin diagrams and the numbering of the positive simple roots.}
\label{dynkin}
\end{figure}

Let $\baser=\{\alpha_1,\ldots,\alpha_N\}$ be the basis of the root system
$\rset$ consisting of positive simple roots. Every positive root
$\beta\in\posr$ has a unique decomposition as a linear combination of elements
of $\baser$ with only positive coefficients. Such basis always exists
\cite{humphreys} and possible bases with these properties are classified by the
corresponding Dynkin diagrams.  Vertices of
such a diagram correspond to simple roots ordered in a particular way,
whereas its edges encode the magnitude of the angle between two consecutive
roots (for details see \cite{humphreys}). We will now restrict our considerations to four
classes of Dynkin diagrams, namely the four so called \emph{classical} infinite
sequences of diagrams of the types $A$, $B$, $C$ and $D$. The numbering of
elements of $\baser$ is shown in the figure \ref{dynkin}. Note that for $B_N$
and $C_N$ algebras $\alpha_1$ is always a long root (the Euclidean length of
roots for a given type of algebra can take at most two values, the roots
with the larger length are called long).

We define the following ordered sets of positive roots (subsets of $\posr$):
\begin{align}\label{ra}
\ra_k:=&\left\{\beta : \beta=\sum_{i=1}^{N} n_i\alpha_i,\,\, n_i=0\mbox{ for }i<k, n_k> 0  \right\},\\
\rb_k:=&\left\{\beta : \beta=\sum_{i=1}^{N} n_i\alpha_i,\,\, n_i=0\mbox{ for }i<k,\,\,\beta\in\posr  \right\}.\label{rb}
\end{align}
For a positive root $\beta=\sum_{i=1}^{N} n_i\alpha_i$, there exists a well
defined notion of the height of the root, $\htr(\beta)=\sum_{i=1}^{N}n_i$. The
sets $\ra_k$ and $\rb_k$ in (\ref{ra}) and (\ref{rb}) are by definition ordered
in the reverse order with respect to the height $\htr(\beta)$ (from the highest
root to the lowest). Observe that
\begin{equation}\label{sumra}
\rb_k=\bigcup_{l=k}^N\ra_l, \qquad \posr=\bigcup_{k=1}^N\ra_k, \qquad
\ra_k\cap\ra_l=\emptyset \mbox{ for } k\neq l.
\end{equation}

Next we define subalgebras generated by corresponding sets of root vectors:
\begin{align}\label{ak}
\ap_k:=&\spn\left\{X_\beta : \beta\in\ra_k  \right\},\nonumber\\
\am_k:=&\spn\left\{X_{-\beta} : \beta\in\ra_k  \right\},\\
\bp_k:=&\bigoplus_{l=k}^{N}\ap_l
  =\spn\left\{X_\beta : \beta\in\rb_k  \right\},\nonumber\\
\bm_k:=&\bigoplus_{l=k}^{N}\am_l
  =\spn\left\{X_{-\beta} : \beta\in\rb_k  \right\}.\label{bk}
\end{align}
The equalities on the right in (\ref{bk}) follow from (\ref{sumra}). The
definitions (\ref{ak}) and (\ref{bk}) and the properties (\ref{commg}) and
(\ref{sumra}) imply that $\ap_k$, $\am_k$, $\bp_k$ and $\bm_k$ are subalgebras
of $\g$ having the following properties:
\begin{align}
\ap_k\subset\bp_k\subset\np&\mbox{ and }
\am_k\subset\bm_k\subset\nm,\label{abprop}\\
\ap_k\cap\ap_l=\emptyset&\mbox{ and }\am_k\cap\am_l=\emptyset \mbox{ for } k\neq l, \nonumber
\end{align}
and we have the following decomposition of a classical Lie algebra $\g$ into a
sum of subalgebras
\begin{equation}\label{commdecomp}
\g=\bigoplus_{i=1}^{N}\ap_i\oplus\h\oplus\bigoplus_{j=1}^{N}\am_j.
\end{equation}
The relation of the definitions (\ref{ak}) and (\ref{bk}) to the diagrams in
Figure~\ref{dynkin} is that $\bp_k$ is the subalgebra of $\np$ generated by the positive
simple roots with first $k-1$ roots from the left side of the diagram omitted.
Observe that for algebras $A_N$, $B_N$ and $D_N$, using the numbering of the
positive simple roots indicated in Figure~\ref{dynkin} and omitting first $k-1$
nodes from the left, one obtains Dynkin diagram corresponding to an algebra of
the type $A_{N-k+1}$, $B_{N-k+1}$ and $D_{N-k+1}$, respectively (for
sufficiently large $N$). This is not the case for $C_N$, where omitting the
first node results in the $A_{N-1}$ Dynkin diagram. This fact distinguishes the
$C_N$ case from the other, but it turns out that the numbering indicated in
Figure~\ref{dynkin} together with the definitions  (\ref{ak}) and (\ref{bk}) is
the unique setting for which the following lemma can be proven. It may be
easily shown by considering explicit counterexamples that for different
numberings of positive simple roots the following will not hold (for example
$\delta=2$ may occur in (\ref{rbform}) and $\bp_k$ may not be commutative).

\begin{Lemma}\label{comma}
The subalgebra $\ap_k$ is a commutative ideal in $\bp_k$ and  $\am_k$ is a
commutative ideal in $\bm_k$, for $k\in\{1,\ldots,N\}$.
\end{Lemma}
\proof Recall that root system $\rset$ is a partially ordered set with respect
to the relation $\prec$ defined as $\alpha\prec\beta$ iff $\beta-\alpha$ is a
positive root or $\beta=\alpha$. Each root system has a unique maximal element
with respect to this relation. These maximal elements have the following forms
in terms of the positive simple roots \cite{humphreys}:
\begin{align}
A_N:&\quad\beta_{max}=\alpha_1+\alpha_2+\ldots+\alpha_N,\nonumber\\
B_N:&\quad\beta_{max}=\alpha_1+2\alpha_2+\ldots+2\alpha_N,\nonumber\\
C_N:&\quad\beta_{max}=\alpha_1+2\alpha_2+\ldots+2\alpha_N,\nonumber\\
D_N:&\quad\beta_{max}=\alpha_1+2\alpha_2+\ldots+2\alpha_{N-2}+\alpha_{N-1}+\alpha_N.
\label{maxroots}
\end{align}
Numbering of simple roots is again as indicated in Figure~\ref{dynkin}. Since
the elements (\ref{maxroots}) are unique and maximal, for any classical algebra
and any positive root $\beta=\sum_{i=1}^{N} n_i\alpha_i$ we have $n_1=0$ or
$n_1=1$ and, according to definition (\ref{ak}), $\ap_1$ is generated by
$X_\beta$ for which $n_1=1$. It follows from (\ref{commg}) that all these
generators commute, since the sum of such roots would have $n_1=2$ and it would
not be a root. So $\ap_1$ is a commutative subalgebra of $\g$. Moreover if
$X_\beta\in\ap_1$ and $X_\gamma\in\bp_1$, then
$[X_\beta,X_\gamma]\subset\spn\left(X_{\beta+\gamma}\right)$ and $\beta+\gamma$
is a root with the first coordinate $n_1=1$ or $[X_\beta,X_\gamma]=0$, so
$[X_\beta,X_\gamma]\in\ap_1$ and $\ap_1$ is a commuting ideal in $\bp_1$.

Consider now $\ap_k$ and $\bp_k$ for $k>1$. In this case we deal with algebras
generated by root vectors $X_\beta$ corresponding to the root system resulting
from omission of first $k-1$ roots (see Figure~ \ref{dynkin}). Thus, the
algebras $\ap_k$ and $\bp_k$ are isomorphic to subalgebras $\ap_1$ and $\bp_1$
of smaller simple algebras of one of the type $A_{N-k+1}$, $B_{N-k+1}$ or
$D_{N-k+1}$ (omission of $\alpha_1$ in $C_N$ yields $A_{N-1}$ root system). It
follows, that $\ap_k$ is a commuting ideal in $\bp_k$.

The proof for subalgebras $\am_k$ and $\bm_k$ is analogous.
\qed

Observe, that the proof of the Lemma \ref{comma} can be summarized in the
following way. If one expresses root $\beta$ in the basis $\baser$ numbered
according to Figure~\ref{dynkin}, $\beta=\sum_{i=1}^{N} n_i\alpha_i$, then
$\beta$ can be identified with the vector of its coefficients:
$\beta=(n_1,n_2,\ldots,n_N)$. In the Lemma \ref{comma} we have shown that in
this notation the functional
$\beta=(\underbrace{0,\ldots,0}_{k-1},2,n_{k+1},\ldots,n_N)$ is never a root
for the classical algebras and the considered simple roots numbering. This fact
together with (\ref{commg}) yields the thesis of the lemma. Thus, it follows
from (\ref{ra}) and the Lemma~\ref{comma} that $\ra_k$ consists of roots of the
following form in the basis $\baser$,
\begin{equation}\label{raform}
\beta=(\underbrace{0,\ldots,0}_{k-1},1,n_{k+1},\ldots,n_N)
\end{equation}
whereas $\rb_k$ consists of roots with coordinates of the form:
\begin{equation}\label{rbform}
\beta=(\underbrace{0,\ldots,0}_{k-1},\delta,n_{k+1},\ldots,n_N),
\end{equation}
where $\delta=0$ or $\delta=1$ and $n_i$ stands for any nonnegative integer
admissible in the given root system.

Since the sets $\ra_k$ defined in (\ref{ra}) are mutually disjoint (see
(\ref{sumra})) this induces an order in the set
\begin{equation}\label{orderphi}
\posr=\bigcup_{k=1}^N\ra_k,
\end{equation}
where we first order elements by the ascending index $k$ corresponding to
$\ra_k$ to which the given element belongs, and then by the order of $\ra_k$.
In what follows we will use the following

\begin{Definition}\label{ordbdef}
The ordered basis $\ordb$ of classical Lie algebra is a basis consisting of
three sequences of generators: first the sequence of root vectors
$\displaystyle\left\{X_\beta: \beta\in\posr\right\}$ ordered in the same order
as the set (\ref{orderphi}), next the sequence of N generators of Cartan
subalgebra $\mathfrak{h}$, and finally the sequence of root vectors
$\displaystyle\left\{X_{-\beta}: \beta\in\posr\right\}$ ordered in the order
reverse to the order of the set (\ref{orderphi}).
\end{Definition}

The main result of this section can be summarized by the following:
\begin{Corollary}\label{cordecomp}
Let $\g$ be a simple complex classical Lie algebra of rank $N$. The equation
(\ref{commdecomp}) defines the decomposition of $\g$ into a sum of $2N+1$
commutative subalgebras.
\end{Corollary}
The basis $\ordb$ is consistent with this splitting of $\g$.

\section{Properties of the adjoint endomorphism}
We now set about formulating and proving the second important ingredient of our
approach concerning the degree of nilpotency and invariance properties of the
adjoint endomorphisms of classical simple Lie algebras.

\begin{Lemma}\label{triang}
Let $X_\alpha\in\g$ be a root vector corresponding to a positive root
$\alpha\in\posr$. In the basis $\ordb$ defined by Definition \ref{ordbdef}, the
matrix of $\ad{X_\alpha}$ is strictly upper triangular and the matrix of
$\ad{X_{-\alpha}}$ is strictly lower triangular.
\end{Lemma}
\proof

Let $\alpha\in\posr$ and $X_\alpha\in\np$. We consider $\ad{X_\alpha}(X)$ for
$X$ being an element of the basis $\ordb$ in three cases:
\begin{enumerate}
\item $X\in\np$, so $X=X_\beta$ for some $\beta\in\posr$. Moreover $X_\alpha\in\ap_k$ and
    $X_\beta\in\ap_l$ for some $k$ and $l$. The roots have the following
    coordinates in terms of the basis $\baser$:
    \begin{equation}\label{triangposr}
    \alpha=(\underbrace{0,\ldots,0}_{k-1},1,n_{k+1},\ldots,n_N), \quad
    \beta=(\underbrace{0,\ldots,0}_{l-1},1,n_{l+1},\ldots,n_N).
    \end{equation}
    If $k=l$ then $\ad{X_\alpha}(X_\beta)=0$.

    For $k<l$, if $\alpha+\beta$ is a root, then $\alpha+\beta\in\ra_k$ and
    $\ad{X_\alpha}(X_\beta)\in\ap_k$, so the matrix element corresponding
    to $\ad{X_\alpha}(X_\beta)$ lies above the diagonal.

    For $k>l$, if $\alpha+\beta$ is a root, then $\alpha+\beta\in\ra_l$ and
    $\ad{X_\alpha}(X_\beta)\in\ap_l$. Since both $\alpha$ and $\beta$ are
    positive roots, the heights fulfill $\htr(\alpha+\beta)>\htr(\beta)$
    hence $X_{\alpha+\beta}$ precedes $X_\beta$ in the basis $\ordb$ and
    the corresponding matrix element lies above the diagonal.
\item $X\in\h$. Since $[\np,\h]\subset\np$, the only nonzero matrix
    elements of $\ad{X_\alpha}$ in the sector corresponding to $\h$ lie
    above the diagonal.
\item $X\in\nm$, so $X=X_\beta$ for $\beta\in\negr$. $X_\alpha\in\ap_k$ and
    $X_\beta\in\am_l$ for some $k$ and $l$. In this case the roots have the
    following coordinates in the basis $\baser$:
    \begin{equation}\label{triangnegr}
    \alpha=(\underbrace{0,\ldots,0}_{k-1},1,n_{k+1},\ldots,n_N), \quad
    \beta=(\underbrace{0,\ldots,0}_{l-1},-1,n_{l+1},\ldots,n_N).
    \end{equation}
    For $k=l$ the first nonzero coordinates in (\ref{triangnegr}) cancel,
    so if $\alpha+\beta$ is a root, then $X_{\alpha+\beta}\in\ap_{l+1}$ or
    $X_{\alpha+\beta}\in\am_{l+1}$. The generators of both $\ap_{l+1}$ and
    $\am_{l+1}$ appear earlier in the basis $\ordb$ than the generators of
    $\am_l$, consequently the matrix element in question lies above the
    diagonal.

    For $k<l$, if $\alpha+\beta$ is a root, then $\alpha+\beta\in\ra_k$ and
    $\ad{X_\alpha}(X_\beta)\in\ap_k$, so the matrix element corresponding
    to $\ad{X_\alpha}(X_\beta)$ lies above the diagonal.

    For $k>l$, if $\alpha+\beta$ is a root, then $\alpha+\beta\in\negr$ and
    $\ad{X_\alpha}(X_\beta)\in\am_l$. Since $\alpha$ is a positive root and
    $\beta$ is a negative one, the heights fulfill
    $\htr(-(\alpha+\beta))<\htr(-\beta)$.  Thus $X_{\alpha+\beta}$ precedes
    $X_\beta$ in the basis $\ordb$ and the corresponding matrix element
    lies above the diagonal.
\end{enumerate}
The statement is also true for $X_\alpha\in\nm$, since it can be obtained by
substituting all the roots by their negatives $\beta\rightarrow-\beta$ and
reversing the order of the basis $\ordb$. \qed

\begin{Lemma}\label{nilpot}
Let $\alpha$ be a root and $X_\alpha\in\g$ be the corresponding root vector
where $\g$ is one of the classical Lie algebras $A_n$, $B_n$, $C_n$ or $D_n$,
then $\left(\ad{X_\alpha}\right)^3=0$.
\end{Lemma}

\proof First observe, that for any $H\in\h$,
\begin{displaymath}
\left(\ad{X_\alpha}\right)^2(H)=[X_\alpha,[X_\alpha,H]]
=-\alpha(H)[X_\alpha,X_\alpha]=0,
\end{displaymath}
thus it suffices to proof that $\left(\ad{X_\alpha}\right)^3(X_\beta)=0$,
$\forall \beta\in\baser$, since the elements $X_\beta$ generate $\np\cup\nm$
and $\g=\np\oplus\h\oplus\nm$. Recall that \cite{humphreys},
\begin{equation}\label{adx1}
\ad{X_\alpha}(X_\beta)=\left[X_\alpha,X_\beta\right]=
\begin{cases}
H_\alpha,&\alpha+\beta=0,\\
N_{\alpha,\beta}X_{\alpha+\beta},&\alpha +\beta\in\baser\\
0,&\mbox{otherwise},
\end{cases}
\end{equation}
where $H_\alpha\in \h$ corresponds to the root $\alpha$ and $N_{\alpha,\beta}$
is some constant. We have
\begin{equation}\label{adx2}
\left(\ad{X_\alpha}\right)^2(X_\beta)=\left[X_\alpha,
\left[X_\alpha,X_\beta\right]\right]=
\begin{cases}
-\alpha(H_\alpha)X_\alpha,&\alpha+\beta=0,\\
N'_{\alpha,\beta}
X_{2\alpha+\beta},&2\alpha +\beta\in\baser,\\
0,&\mbox{otherwise},
\end{cases}
\end{equation}
where $N'_{\alpha,\beta}$ is some other constant. Finally, since
$\left[X_\alpha,X_\alpha\right]=0$ we obtain:
\begin{equation}\label{adx3}
\left(\ad{X_\alpha}\right)^3(X_\beta)=\left[X_\alpha,\left[X_\alpha,
\left[X_\alpha,X_\beta\right]\right]\right]=
\begin{cases}
N''_{\alpha,\beta}
X_{3\alpha+\beta},&3\alpha +\beta\in\baser,\\
0,&\mbox{otherwise.}
\end{cases}
\end{equation}
We will show that $3\alpha+\beta$ is never a root for a classical algebra $\g$.
A pair of roots $\alpha,\beta$ generates a 2-dimensional root system. According
to \cite{humphreys} there are exactly four 2-dimensional root systems:
$A_1\times A_1$, $A_2$, $B_3$ and $G_2$. The equations (\ref{adx1}-\ref{adx3})
imply that $\left(\ad{X_\alpha}\right)^3(X_\beta)\neq0$, provided there exist
a, so called, root string of length four consisting of roots $\beta$,
$\alpha+\beta$, $2\alpha+\beta$ and $3\alpha+\beta$. But the only 2-dimensional
root system containing a root string of length four is the $G_2$ root system
and the angle between roots $\alpha$ and $\beta$ is equal $150^\circ$ in this
case. On the other hand according to classification of simple Lee algebras
\cite{humphreys}, the only simple algebra with a pair of roots connected by
such an angle is the $G_2$ algebra itself. So no other simple Lie algebra can
have root system containing root string or length four. \qed

Observe that if $\alpha\in\ra$ and $\beta\in\ra$ then $\ad{X_\alpha}$ and
$\ad{X_\beta}$ are commuting nilpotent operators of nilpotency order 3. In
general, if we have two commuting matrices of a given nilpotency order $r$ then
the sum of the matrices is also nilpotent of order $r$. It follows from the
Jordan theorem \cite{humphreys} - the matrices can be expressed in the Jordan
form in the same basis and are block-diagonal with the same blocks of maximal
size $r-1$. So the Lemmas \ref{comma} and \ref{nilpot} yield:
\begin{Corollary}\label{anilpot}
If $X\in\ap_k$ or $X\in\am_k$ then $(\ad{X})^3=0$.
\end{Corollary}
Next we proof the crucial invariance property of the decomposition
(\ref{commdecomp}).
\begin{Lemma}\label{invsp}
Let $X_\alpha\in\ap_k\subset\g$ or
$X_\alpha\in\am_k\subset\g$, where $\alpha\in\ra$ is the corresponding
root. The subalgebras $\ap_l$, $\am_l$ for $l<k$ and the subalgebra
$\bp_k\oplus\h\oplus\bm_k$ are invariant subspaces of
$\ad{X_\alpha}$.
\end{Lemma}
\proof Let $X_\alpha\in\ap_k$. We consider three cases:
\begin{enumerate}
\item $X_\beta\in\ap_l$, $l<k$. It follows from  (\ref{ak}) and
    (\ref{abprop}) that $\ap_k\subset\bp_k\subset\bp_l$. In this case
    $X_\alpha\in\bp_l$ and $X_\beta\in\bp_l$. Thus,
    $\ad{X_\alpha}(X_\beta)=[X_\alpha,X_\beta]\in\bp_l$. On the other hand
    $X_\beta\in\ap_l$ and by Lemma \ref{comma},  $\ap_l$ is an ideal in
    $\bp_l$, so $\ad{X_\alpha}(X_\beta)\in\ap_l$.
\item $Y\in\bp_k\oplus\h\oplus\bm_k$. We have also
    $X_\alpha\in\bp_k\oplus\h\oplus\bm_k$ and the property
    $\ad{X_\alpha}(Y)\in\bp_k\oplus\h\oplus\bm_k$ follows from the fact
    that $\bp_k\oplus\h\oplus\bm_k$ is subalgebra of $\g$ and this a direct
    consequence of the definition of $\bp_k$ and $\bm_k$ (see (\ref{bk})).
\item $X_\beta\in\am_l$, $l<k$. Since $\alpha\in\ra_k$ and
    $-\beta\in\rb_l$, in this case (see definitions (\ref{ra}-\ref{rb}))
    the coordinates of roots $\alpha$ and $\beta$ in the basis $\baser$
    have the following form (see (\ref{raform}-\ref{rbform})):
\begin{equation}
\alpha=(\underbrace{0,\ldots,0}_{k-1},1,n_{k+1},\ldots,n_N),\quad
\beta=(\underbrace{0,\ldots,0}_{l-1},-1,n_{l+1},\ldots,n_N),
\end{equation}
Since $l<k$, the sum $\alpha+\beta$ has the form:
\begin{equation}\label{sumrootsform}
\alpha+\beta=(\underbrace{0,\ldots,0}_{l-1},-1,n_{l+1},\ldots,n_N),
\end{equation}
so if $\alpha+\beta$ is not a root then $\ad{X_\alpha}(X_\beta)=0$ and if
$\alpha+\beta$ is a root then since it has the form (\ref{sumrootsform}),
we have $\alpha+\beta\in\ra_l$ and
$\ad{X_\alpha}(X_\beta)=X_{\alpha+\beta}\in\ap_l$.
\end{enumerate}
The analogous reasoning holds for $X_\alpha\in\am_k$.\qed

\section{Exponential mapping}
The lemmas proved in the preceding sections allow now to exhibit the structure
of the exponential of the adjoint endomorphism. Its quadrating dependence on
the exponent results in the Wei-Norman equations in the Riccati form, whereas
its triangular and block-diagonal structure orders the resulting Riccati
equations in a specific hierarchy.

Lemma \ref{invsp} implies that $\ad{X}$ for $X\in\ap_k$ or $X\in\am_k$ is a
block diagonal operator with respect to the following decomposition:

\begin{equation}\label{blocks}
\g=\ap_1\oplus\ldots\oplus\ap_{k-1}\oplus
\underbrace{\left(
\bp_k\oplus\h\oplus\bm_k
\right)}_{\mbox{one block}}
\oplus\am_{k-1}\oplus\ldots\oplus\am_1.
\end{equation}
From Corollary \ref{anilpot} we know that it is also nilpotent, $\ad{X}^3=0$.
Thus Corollary \ref{anilpot} and Lemmas \ref{invsp} and \ref{triang} yield:

\begin{Corollary}\label{expprop}
For $X\in\ap_k$ or $X\in\am_k$ the matrix of $\exp(\ad{X})$ is a quadratic
polynomial in $\ad{X}$ and it is block diagonal with respect to the
decomposition (\ref{blocks}). Moreover in the basis $\ordb$ (Definition
\ref{ordbdef}) the matrix of $\exp(\ad{X})$ is upper triangular for $X\in\ap_k$
and lower triangular for $X\in\am_k$.
\end{Corollary}

The basis $\ordb$ defined by Definition \ref{ordbdef} is adjusted to the
decomposition (\ref{blocks}) in the sense that successive blocks in the
decomposition (\ref{blocks}) are generated by successive subsequences of the
basis $\ordb$. It is thus convenient to rewrite the equation (\ref{eq-wn0}) as
a sum of terms corresponding to the decomposition (\ref{blocks}). We denote
elements of $\ordb$ by $X_i$ and define:
\begin{equation}\label{defU}
\up_k:=\prod_{X_i\in \ap_k} \exp(u_i\ad{X_i}),
 \qquad
 \um_k:=\prod_{X_i\in \am_k} \exp(u_i\ad{X_i}).
\end{equation}
Since $\ap_k$ and $\am_k$ are commutative subalgebras, the order of factors in
(\ref{defU}) does not matter. For $l<\dim\ap_k$ we also define:
\begin{equation}\label{defV}
\vp_{kl}:=\prod_{i\in I_{kl}} \exp(u_i\ad{X_i}),
 \qquad
 \vm_{kl}:=\prod_{i\in \tilde{I}_{kl}} \exp(u_i\ad{X_i}),
\end{equation}
where for a given $k$ the index $i$ runs over the first $l-1$ generators of
$\ap_k$ and $\am_k$ respectively. So $I_{kl}$ is the string of the first (with
respect to the ordering of $\ordb$) $l$ indices $i$, such that $X_i\in\ap_k$,
and $X_i\in\ordb$ and analogously, $I_{kl}$ is the sequence of the first $l$
indices $i$, such that $X_i\in\am_k$, and $X_i\in\ordb$.  We set
$$
\up_0=\um_N=\vp_{k0}=\vm_{k0}=\mathbbm{1}.
$$
Observe that commutativity of the subalgebras $\ap_k$ implies that:
\begin{displaymath}
\up_k=\prod_{X_i\in \am_k} \exp(u_i\ad{X_i})
 =\exp\left(\sum_{X_i\in \am_k} u_i\ad{X_i} \right)
 =\exp\left(\ad{\left(\sum_{X_i\in \am_k} u_iX_i\right)} \right),
\end{displaymath}
hence $\up_k$ equals $\exp(\ad{X})$ for some $X\in\ap_k$ (analogously $\um_k$
equals $\exp(\ad{X})$ for some $X\in\am_k$). Thus by Corollary \ref{expprop} we
have:
\begin{Corollary}\label{quadratic}
The matrix elements of the operators $\up_k$, $\vp_{kl}$, $\um_k$ and
$\vm_{kl}$ are polynomial functions of the parameters $u_i$ of degree at most
2. The matrices $\up_k$, $\vp_{kl}$, $\um_k$ and $\vm_{kl}$ are block diagonal
with respect to the decomposition (\ref{blocks}) and the matrices $\up_k$ and
$\vp_{kl}$ are upper triangular, whereas the matrices $\um_k$ and $\vm_{kl}$
are lower triangular.
\end{Corollary}

We also define exponentials corresponding to Cartan subalgebra $\h$:
\begin{equation}\label{defH}
H:=\prod_{X_i\in\h}\exp(u_i\ad{X_i}),
 \qquad
 H_l:=\prod_{i\in I^\h_l}\exp(u_i\ad{X_i}),
\end{equation}
where $I^\h_l$ is the index range numbering the first (with respect to the
ordering of $\ordb$) $l$ generators of $\h$. $H$ and $H_l$ are diagonal
matrices and we set $H_0=\mathbbm{1}$.

\section{Wei-Norman method and example results}
\label{secresults}

Since the operators defined in (\ref{defU}), (\ref{defV}) and (\ref{defH}) have
exactly the same properties as corresponding operators in Section 5 of
\cite{charzynski13}, the method of separation of equation (\ref{eq-wn0}) into
subsystems corresponding to the elements of the decomposition
(\ref{commdecomp}) described in Section 6 of \cite{charzynski13} works without
any modification. We will not repeat the proof given in \cite{charzynski13}
here, but we will show a few example results for low values of $N$. It should
be stressed that a crucial ingredient for realizing the described in
\cite{charzynski13} algorithm in practice is the order of the generators in the
basis $\ordb$ described in Definition~\ref{ordbdef}. Once this basis is used
for computations and the inverse in (\ref{eq-wn3}) is successfully computed,
the separation of the system of equations comes up automatically.

In what follows we present example results. The results for $A_N$, $N=1, 2, 3$
where presented in \cite{charzynski13}. Here we present the examples not
equivalent to those presented already in \cite{charzynski13} and not equivalent
to each other, namely: $A_4$, $B_2,$ $B_3$, $B_4$, $C_3$, $C_4$ and $D_4$.
(There are the so called \emph{accidental isomorphisms} in low dimensions:
 $A_1\simeq B_1 \simeq C_1$, $D_2\simeq A_1\times A_1$, $B_2\simeq C_2$ and $A_3\simeq D_3$.)
In what follows we use matrix representation of classical algebras as traceless
matrices preserving given bilinear form. There are many equivalent conventions
in defining these preserved bilinear forms. We use the same convention as in
\cite{humphreys}. In all cases we provide explicit parametrization of the
considered matrix algebra. The results for $B_2$ case are presented in great detail. For all examples we give the basis $\ordb$ fulfilling the Definition~\ref{ordbdef}, which is crucial for the calculations. Once the basis is known, the equations for $u_i$ can be easily computed, so we do not present them for $N=4$ and for $N=3$ we give only the Riccati equations corresponding to $\np$.

\subsection{$B_2$ algebra (Lie algebra of $O(5,\C)$ group and $Sp(4,\C)$)}

The ordered basis of $B_2$ algebra fulfilling the Definition \ref{ordbdef} is
encoded in the matrix:
\begin{equation}\label{b2base}
\sum_{i=1}^{10}a_iX_i=
\left[ \begin {array}{ccccc} 0&-a_{{9}}&-a_{{4}}&a_{{2}}&a_{{7}}
\\-a_{{2}}&a_{{5}}&a_{{1}}&0&a_{{3}}
\\-a_{{7}}&a_{{10}}&a_{{5}}-2\,a_{{6}}&-a_{{3}}&0
\\a_{{9}}&0&-a_{{8}}&-a_{{5}}&-a_{{10}}
\\a_{{4}}&a_{{8}}&0&-a_{{1}}&2\,a_{{6}}-a_{{5}}
\end {array} \right].
\end{equation}
According to (\ref{commdecomp}) the algebra of type $B_2$ splits in the
following way:
\begin{equation}\label{b2split}
\g=\ap_1\oplus\ap_2\oplus\h\oplus\am_2\oplus\am_1.
\end{equation}
The order of the basis defined in (\ref{b2base}) is consistent with the
splitting (\ref{b2split}) in the sense that
 $\ap_1=\spn\left\{X_1,X_2,X_3\right\}$,
 $\ap_2=\spn\left\{X_4\right\}$,
 $\h=\spn\left\{X_5,X_6\right\}$,
 $\am_2=\spn\left\{X_7\right\}$ and
 $\am_1=\spn\left\{X_8,X_9,X_{10}\right\}$. The decomposition (\ref{b2split})
yields the separation of system (\ref{eq-wn3}) into five subsystems. First we
get the $3$-dimensional matrix Riccati equation corresponding to $\ap_1$:
\begin{align*}
u^{\prime}_{{1}}=&
 -a_{{10}}{u_{{1}}}^{2}
 +\frac12\,a_{{8}}{u_{{2}}}^{2}
 -a_{{9}}u_{{1}}u_{{2}}
 -a_{{4}}u_{{2}}+2\,a_{{6}}u_{{1}}+a_{{1}},
\\
u^{\prime}_{{2}}=&
 -\frac12\,a_{{9}}{u_{{2}}}^{2}
 -a_{{10}}u_{{1}}u_{{2}}
 -a_{{8}}u_{{2}}u_{{3}}
 +a_{{9}}u_{{1}}u_{{3}}
 +a_{{4}}u_{{3}}
 +a_{{5}}u_{{2}}
 -a_{{7}}u_{{1}}+a_{{2}},
\\
u^{\prime}_{{3}}=&
 \frac12\,a_{{10}}{u_{{2}}}^{2}
 -a_{{8}}{u_{{3}}}^{2}
 -a_{{9}}u_{{2}}u_{{3}}
 +a_{{7}}u_{{2}}
 +\left( 2\,a_{{5}}-2\,a_{{6}} \right)u_{{3}}
 +a_{{3}}.
\end{align*}
for $u_1$, $u_2$ and $u_3$. Once we solve it, we can solve the Riccati equation
for $u_4$:
\begin{align*}
u^{\prime}_{{4}}=&
 \frac12\left(a_{{9}}u_{{3}}-a_{{10}}u_{{2}}-a_{{7}} \right) {u_{{4}}}^{2}
 +\left( a_{{8}}u_{{3}}-a_{{10}}u_{{1}}-a_{{5}}+2\,a_{{6}} \right) u_{{4}}-a_{{8}}u_{{2}}+a_{{9}}u_{{1}}
 +a_{{4}}.
\end{align*}
After the solution of this equation is found the rest of the solutions for
$u_i$ can be found by simple consecutive integrations. The equations
corresponding to $h$ are:
\begin{align*}
u^{\prime}_{{5}}=&-a_{{8}}u_{{3}}-a_{{9}}u_{{2}}-a_{{10}}u_{{1}}+a_{{5}},
\\
u^{\prime}_{{6}}=&\frac12\,a_{{9}}u_{{3}}u_{{4}}-\frac12\,a_{{10}}u_{{2}}u_{{4}}-\frac12\,a_
{{7}}u_{{4}}-\frac12\,a_{{9}}u_{{2}}-a_{{10}}u_{{1}}+a_{{6}}.
\end{align*}
The equation corresponding to the sector $\am_2$ reads
\begin{displaymath}
u^{\prime}_{{7}}=- \left( a_{{9}}u_{{3}}-a_{{10}}u_{{2}}-a_{{7}} \right) { {\rm
e}^{-u_{{5}}+2\,u_{{6}}}}\, .
\end{displaymath}
And finally, the equations for $\am_1$:
\begin{align*}
u^{\prime}_{{8}}=& \left( -\frac12\,a_{{10}}{u_{{4}}}^{2}+a_{{9}}u_{{4}}+a_{{8}}
 \right) {{\rm e}^{2\,u_{{5}}-2\,u_{{6}}}},
\\
u^{\prime}_{{9}}=&\frac12\, \left( -a_{{10}}{u_{{4}}}^{2}+2\,a_{{9}}u_{{4}}+2\,a_{{8
}} \right) u_{{7}}{{\rm e}^{2\,u_{{5}}-2\,u_{{6}}}}+ \left( -a_{{10}}u
_{{4}}+a_{{9}} \right) {{\rm e}^{u_{{5}}}},
\\
u^{\prime}_{{10}}=&-\frac14\, \left( -a_{{10}}{u_{{4}}}^{2}+2\,a_{{9}}u_{{4}}+2\,a_{
{8}} \right) {u_{{7}}}^{2}{{\rm e}^{2\,u_{{5}}-2\,u_{{6}}}}+
a_{{10}}{
{\rm e}^{2\,u_{{6}}}}- \left( -a_{{10}}u_{{4}}+a_{{9}} \right) u_{{7}}
{{\rm e}^{u_{{5}}}}.
\end{align*}

\subsection{$B_3$ algebra (Lie algebra of $O(7,\C)$ group and $Sp(6,\C)$)}

In this case the algebra decomposes into $\g=\ap_1\oplus\ap_2\oplus\ap_3\oplus
\h\oplus\am_3\oplus\am_2\oplus\am_1$ and the ordered basis fulfilling the
Definition~\ref{ordbdef} yields the following parametrization:
\begin{displaymath}
\sum_{i=1}^{21}a_iX_i=
 \left[ \begin {array}{ccccccc} 0&-a_{{19}}&-a_{{7}}&-a_{{9}}&a_{{3}}&
a_{{15}}&a_{{13}}\\-a_{{3}}&a_{{10}}&a_{{1}}&a_{{2}}
&0&a_{{5}}&a_{{4}}\\-a_{{15}}&a_{{21}}&a_{{10}}-a_{{
11}}&a_{{14}}&-a_{{5}}&0&a_{{16}}\\-a_{{13}}&a_{{20}
}&a_{{8}}&a_{{11}}-2\,a_{{12}}&-a_{{4}}&-a_{{16}}&0
\\a_{{19}}&0&-a_{{17}}&-a_{{18}}&-a_{{10}}&-a_{{21}}
&-a_{{20}}\\a_{{7}}&a_{{17}}&0&-a_{{6}}&-a_{{1}}&-a_
{{10}}+a_{{11}}&-a_{{8}}\\a_{{9}}&a_{{18}}&a_{{6}}&0
&-a_{{2}}&-a_{{14}}&-a_{{11}}+2\,a_{{12}}\end {array} \right],
\end{displaymath}
where
 $\ap_1=\spn\left\{X_1,X_2,X_3,X_4,X_5\right\}$,
 $\ap_2=\spn\left\{X_6,X_7,X_8\right\}$,
 $\ap_3=\spn\left\{X_9\right\}$,\\
 $\h=\spn\left\{X_{10},X_{11},X_{12}\right\}$,
 $\am_3=\spn\left\{X_{13}\right\}$,
 $\am_2=\spn\left\{X_{14},X_{15},X_{16}\right\}$\\ and
 $\am_1=\spn\left\{X_{17},X_{18},X_{19},X_{20},X_{21}\right\}$.

The matrix Riccati equation corresponding to $\ap_1$ reads:
\begin{align*}
u^{\prime}_{{1}}=&a_{{17}}u_{{2}}u_{{4}}+\frac12\,a_{{17}}{u_{{3}}}^{2}
 -a_{{18}}u_{{1}}u_{{4}}-a_{{19}}u_{{1}}u_{{3}}
 -a_{{20}}u_{{1}}u_{{2}}-a_{{21}}{u_{{1}}}^{2}+\\
 &-a_{{6}}u_{{4}}-a_{{7}}u_{{3}}-a_{{8}}u_{{2}}+a_{{11}}u_{{1}}+
a_{{1}},
\\
u^{\prime}_{{2}}=&-a_{{17}}u_{{2}}u_{{5}}+a_{{18}}u_{{1}}u_{{5}}+
 \frac12\,a_{{18}}{u_{{3}}}^{2}-a_{{19}}u_{{2}}u_{{3}}
 -a_{{20}}{u_{{2}}}^{2}-a_{{21}}u_{{1}}u_{{2}}+\\
 & +a_{{6}}u_{{5}}-a_{{9}}u_{{3}}-a_{{14}}u_{{1}}+
 \left( a_{{10}}-a_{{11}}+2\,a_{{12}} \right) u_{{2}}+a_{{2}},
\\
u^{\prime}_{{3}}=&-a_{{17}}u_{{3}}u_{{5}}-a_{{18}}u_{{3}}u_{{4}}
 +a_{{19}}u_{{1}}u_{{5}}+a_{{19}}u_{{2}}u_{{4}}
 -\frac12\,a_{{19}}{u_{{3}}}^{2}-a_{{20}}u_{{2}}u_{{3}}-a_{{21}}u_{{1}}u_{{3}}+\\
 & +a_{{7}}u_{{5}}+a_{{9}}u_{{4}}
 +a_{{10}}u_{{3}}-a_{{13}}u_{{2}}-a_{{15}}u_{{1}}+a_{{3}},
\\
u^{\prime}_{{4}}=&\frac12\,{u_{{3}}}^{2}a_{{20}}-a_{{19}}u_{{3}}u_{{4}}
 -a_{{18}}{u_{{4}}}^{2}+a_{{20}}u_{{5}}u_{{1}}
 -a_{{21}}u_{{1}}u_{{4}}-a_{{17}}u_{{5}}u_{{4}}+\\
 &+u_{{3}}a_{{13}}-a_{{16}}u_{{1}}+a_{{8}}u_{{5}}-
 \left( 2\,a_{{12}}-a_{{10}}-a_{{11}} \right) u_{{4}}+a_{{4}}
,
\\
u^{\prime}_{{5}}=&-a_{{17}}{u_{{5}}}^{2}-a_{{18}}u_{{4}}u_{{5}}
 -a_{{19}}u_{{3}}u_{{5}}-a_{{20}}u_{{2}}u_{{5}}
 +a_{{21}}u_{{2}}u_{{4}}+\frac12\,a_{{21}}{u_{{3}}}^{2}+\\
 &+a_{{14}}u_{{4}}+a_{{15}}u_{{3}}+a_{{16}}u_{{2}}+
 \left( 2\,a_{{10}}-a_{{11}} \right) u_{{5}}+a_{{5}}.
\end{align*}
The system of equation corresponding to $\ap_2$ is
\begin{align*}
u^{\prime}_{{6}}=& \left( a_{{20}}u_{{5}}-a_{{21}}u_{{4}}
 -a_{{16}} \right) {u_{{6}}}^{2}+ \left( a_{{19}}u_{{5}}
 -a_{{21}}u_{{3}}-a_{{15}} \right) u_{{6}}u_{{7}}+\\
 &+ \frac12\left( -a_{{18}}u_{{5}}+a_{{21}}u_{{2}} +a_{{14}} \right) {u_{{7}}}^{2}
 + \left( a_{{18}}u_{{3}}-a_{{19}}u_{{2}}-a_{{9}} \right) u_{{7}}+\\
 & + \left( a_{{17}}u_{{5}}+a_{{18}}u_{{4}}-a_{{20}}u_{{2}}-a_{{21}}u_{{1}}-a_{{10}}+2\,a_{{12}} \right) u_{{6}}
 -a_{{17}}u_{{2}}+a_{{18}}u_{{1}}+a_{{6}},
\\
u^{\prime}_{{7}}=&
 \left(a_{{20}}u_{{5}}-a_{{ 21}}u_{{4}}-a_{{16}} \right) u_{{6}}u_{{7}}
 +\left( u_{{3}}a_{{21}}-a_ {{19}}u_{{5}}+a_{{15}} \right)u_{{6}}u_{{8}}+\\
 &-\frac12\, \left( u_{{3}}a_{{21}}-a_{{19}}u_{{5}}+a_{{15}} \right) {u_{{7}}}^{2}
 - \left(u_{{2}}a_{ {21}}-a_{{18}}u_{{5}}+a_{{14}} \right)u_{{7}}u_{{8}}+\\
 &- \left(u_{{3}}a_{{20}}-a_{{19}}u_{{4}}+a_{{13}} \right) u_{{6}}
 - \left(a_{{ 21}}u_{{1}}-a_{{17}}u_{{5}}+a_{{10}}-a_{{11}} \right) u_{{7}}+\\
 &-\left( u_{{3}}a_{{18}}-u_{{2}}a_{{19}}-a_{{9}} \right) u_{{8}}
 -u_{{3}}a_{{17}}
 +a_{{19}}u_{{1}}
 +a_{{7}},
\\
u^{\prime}_{{8}}=&
 -\frac12\, \left(a_{{20}}u_{{5}}-a _{{21}}u_{{4}}-a_{{16}} \right) {u_{{7}}}^{2}
 -\left( u_{{3}}a_{{21}}- a_{{19}}u_{{5}}+a_{{15}} \right)u_{{7}}u_{{8}}+\\
 &- \left( u_{{2}}a_{{21} }-a_{{18}}u_{{5}}+a_{{14}}\right) {u_{{8}}}^{2}
 + \left( u_{{3}}a_{{20}}-a_{{19}}u_{{4}}+a_{{13}} \right) u_{{7}}+\\
 &+ \left(u_{{2}}a_{{20}}-a_{{18}}u_{{4}}-a_{{21}}u_{{1}}+a_{{17}}u_{{5}}-2\,a_{{12}}-
a_{{10}}+2\,a_{{11}} \right) u_{{8}}+a_{{20}}u_{{1}}-a_{{17}}u_{{4}}
 +a_{{8}}.
\end{align*}
And the last equation for $\ap_3$ reads
\begin{align*}
u^{\prime}_{{9}}=&
 -\frac12\, \Big(-u_{{3}}a_{{21}}u_{{8}}+a_{{19}}u_{{5}}u_{{8}}
 -a_{{20}}u_{{5}}u_{{7}}+a_{{21}}u_{{7}}u_{{4}}+\\
 &\hspace{2cm}+a_{{20}}u_{{3}}-a_{{19}}u_{{4}}
 -a_{{15}}u_{{8}}+a_{{16}}u_{{7}}+a_{{13}} \Big) {u_{{9}}}^{2}
 +\\ &
 -\Big( a_{{18}}u_{{5}}u_{{8}}-a_{{21}}u_{{2}}u_{{8}}-a_{{20}}u_{{6}}u_{{5}}
 +a_{{21}}u_{{6}}u_{{4}}
 +\\ &\hspace{1cm}
 +u_{{2}}a_{{20}}-a_{{18}}u_{{4}}-a_{{14}}u_{{8}}
 +a_{{16}}u_{{6}}-2\,a_{{12}}+a_{{11}} \Big) u_{{9}}+\\
 &+a_{{21}}u_{{3}}u_{{6}}-a_{{21}}u_{{2}}u_{{7}}+a_{{18}}u_{{5}}u_{{7}}-a_{{19}}u_{{6}}u_{{5}}
 -a_{{18}}u_{{3}}+a_{{19}}u_{{2}}-a_{{14}}u_{{7}}+a_{{15}}u_{{6}}
 +a_{{9}}.
\end{align*}
The remaining 12 equations which we do not present here can be solved by
consecutive integrations.

\subsection{$C_3$ algebra (Lie algebra of $Sp(6,\C)$ group)}

In this case the ordered basis fulfilling the Definition~\ref{ordbdef} yields
the following parametrization:

\begin{displaymath}
\sum_{i=1}^{21}a_iX_i=
\left[ \begin {array}{cccccc}
a_{{10}}-a_{{11}}&a_{{14}}&a_{{15}}&a_{{6}}&a_{{5}}&a_{{3}}\\a_{{8}}&a_{{11}}-a_{{12}}&a_{{
13}}&a_{{5}}&a_{{4}}&a_{{2}}\\a_{{7}}&a_{{9}}&a_{{12
}}&a_{{3}}&a_{{2}}&a_{{1}}\\a_{{16}}&a_{{17}}&a_{{19
}}&-a_{{10}}+a_{{11}}&-a_{{8}}&-a_{{7}}\\a_{{17}}&a_
{{18}}&a_{{20}}&-a_{{14}}&a_{{12}}-a_{{11}}&-a_{{9}}
\\a_{{19}}&a_{{20}}&a_{{21}}&-a_{{15}}&-a_{{13}}&-a_
{{12}}\end {array} \right],
\end{displaymath}
which is consistent with the following decomposition:
\begin{align*}
\g=&\ap_1\oplus\ap_2\oplus\ap_3\oplus
\h\oplus\am_3\oplus\am_2\oplus\am_1=\\
 =&\spn\left\{X_1,\ldots,X_6\right\}
 \oplus\spn\left\{X_7,X_8\right\}
 \oplus\spn\left\{X_9\right\}\oplus\\
 &\oplus\spn\left\{X_{10},X_{11},X_{12}\right\}\oplus\\
 &\oplus\spn\left\{X_{13}\right\}
 \oplus\spn\left\{X_{14},X_{15}\right\}
 \oplus\spn\left\{X_{16},\ldots,X_{21}\right\}.
\end{align*}
The equations corresponding to $\ap_1$, $\ap_2$ and $\ap_3$ are:
\begin{align*}
u^{\prime}_{{1}}=&
 -a_{{16}}{u_{{3}}}^{2}-2\,a_{{17}}u_{{2}}u_{{3}}
 -a_{{18}}{u_{{2}}}^{2}-2\,a_{{19}}u_{{1}}u_{{3}}
 -2\,a_{{20}}u_{{1}}u_{{2}}-a_{{21}}{u_{{1}}}^{2}+\\
 & +2\,a_{{7}}u_{{3}}+2\,a_{{9}}u_{{2}}
 +2\,a_{{12}}u_{{1}}+a_{{1}},
\\
u^{\prime}_{{2}}=&
 -a_{{16}}u_{{3}}u_{{5}}-a_{{17}}u_{{2}}u_{{5}}
  -a_{{17}}u_{{3}}u_{{4}}-a_{{18}}u_{{2}}u_{{4}}
  -a_{{19}}u_{{1}}u_{{5}}-a_{{19}}u_{{2}}u_{{3}}+\\
  &-a_{{20}}u_{{1}}u_{{4}}-a_{{20}}{u_{{2}}}^{2}
  -a_{{21}}u_{{1}}u_{{2}}+a_{{7}}u_{{5}}
  +a_{{8}}u_{{3}}+a_{{9}}u_{{4}}
  +a_{{11}}u_{{2}}+a_{{13}}u_{{1}}+a_{{2}},
\\
u^{\prime}_{{3}}=&
 -a_{{16}}u_{{3}}u_{{6}}-a_{{17}}u_{{2}}u_{{6}}
 -a_{{17}}u_{{3}}u_{{5}}-a_{{18}}u_{{2}}u_{{5}}
 -a_{{19}}u_{{1}}u_{{6}}-a_{{19}}{u_{{3}}}^{2}+\\
 &
 -a_{{20}}u_{{1}}u_{{5}}-a_{{20}}u_{{2}}u_{{3}}
 -a_{{21}}u_{{1}}u_{{3}}+\\
 &+a_{{7}}u_{{6}}
 +a_{{9}}u_{{5}}+a_{{14}}u_{{2}}
 +a_{{15}}u_{{1}}
 + \left( a_{{10}}-a_{{11}}+a_{{12}} \right) u_{{3}}+a_{{3}},
\\
u^{\prime}_{{4}}=&
 -a_{{16}}{u_{{5}}}^{2}-2\,a_{{17}}u_{{4}}u_{{5}}
 -a_{{18}}{u_{{4}}}^{2}-2\,a_{{19}}u_{{2}}u_{{5}}
 -2\,a_{{20}}u_{{2}}u_{{4}}-a_{{21}}{u_{{2}}}^{2}+\\
 &+2\,a_{{8}}u_{{5}}+2\,a_{{13}}u_{{2}}
 + \left( 2\,a_{{11}}-2\,a_{{12}} \right) u_{{4}}+a_{{4}},
\\
u^{\prime}_{{5}}=&
 -a_{{16}}u_{{5}}u_{{6}}-a_{{17}}u_{{4}}u_{{6}}
 -a_{{17}}{u_{{5}}}^{2}-a_{{18}}u_{{4}}u_{{5}}
 -a_{{19}}u_{{2}}u_{{6}}-a_{{19}}u_{{3}}u_{{5}}
 -a_{{20}}u_{{2}}u_{{5}}+\\
 &-a_{{20}}u_{{3}}u_{{4}}
 -a_{{21}}u_{{2}}u_{{3}}+
 a_{{8}}u_{{6}}
 +a_{{13}}u_{{3}}+a_{{14}}u_{{4}}
 +a_{{15}}u_{{2}}
 + \left( a_{{10}}-a_{{12}} \right) u_{{5}}+a_{{5}},
\\
u^{\prime}_{{6}}=&
 -a_{{16}}{u_{{6}}}^{2}-2\,a_{{17}}u_{{5}}u_{{6}}
 -a_{{18}}{u_{{5}}}^{2}-2\,a_{{19}}u_{{3}}u_{{6}}
 -2\,a_{{20}}u_{{3}}u_{{5}}-a_{{21}}{u_{{3}}}^{2}+\\
 &
 +2\,a_{{14}}u_{{5}}+2\,a_{{15}}u_{{3}}
+ \left( 2\,a_{{10}}-2\,a_{{11}} \right) u_{{6}} +a_{{6}},
\\
u^{\prime}_{{7}}=&
 + \left( a_{{19}}u_{{6}}+a_{{20}}u_{{5}}+a_{{21}}u_{{3}}-a_{{15}} \right) {u_{{7}}}^{2}
 + \left( a_{{17}}u_{{6}}+a_{{18}}u_{{5}}+a_{{20}}u_{{3}}-a_{{14}} \right) u_{{7}}u_{{8}}+\\
 &
 + \left( a_{{16}}u_{{6}}+a_{{17}}u_{{5}}-a_{{20}}u_{{2}}
     -a_{{21}}u_{{1}}-a_{{10}}+a_{{11}}+a_{{12}} \right) u_{{7}}+\\
 &
 + \left( -a_{{17}}u_{{3}}-a_{{18}}u_{{2}}-a_{{20}}u_{{1}}+a_{{9}} \right) u_{{8}}
 -a_{{16}}u_{{3}}-a_{{17}}u_{{2}}-a_{{19}}u_{{1}}
 +a_{{7}},
\end{align*}
\begin{align*}
u^{\prime}_{{8}}=&
 + \left( a_{{19}}u_{{6}}+a_{{20}}u_{{5}}+a_{{21}}u_{{3}}-a_{{15}} \right) u_{{7}}u_{{8}}
 + \left( a_{{17}}u_{{6}}+a_{{18}}u_{{5}}+a_{{20}}u_{{3}}-a_{{14}}
         \right) {u_{{8}}}^{2}+\\
 &
 + \left( a_{{16}}u_{{6}}-a_{{18}}u_{{4}}+a_{{19}}u_{{3}}
       -a_{{20}}u_{{2}}-a_{{10}}+2\,a_{{11}}-a_{{12}} \right) u_{{8}}+\\
 &
 + \left( -a_{{19}}u_{{5}}-a_{{20}}u_{{4}}-a_{{21}}u_{{2}}+a_{{13}} \right) u_{{7}}
 -a_{{16}}u_{{5}}-a_{{17}}u_{{4}}-a_{{19}}u_{{2}}+a_{{8}},
\\
u^{\prime}_{{9}}=&
  \left( -a_{{19}}u_{{6}}u_{{8}}-a_{{20}}u_{{5}}u_{{8}}
   -a_{{21}}u_{{3}}u_{{8}}+a_{{15}}u_{{8}}
 +a_{{19}}u_{{5}}
   +a_{{20}}u_{{4}}+a_{{21}}u_{{2}}-a_{{13}} \right) {u_{{9}}}^{2}+\\
 &
  +\Big( -a_{{17}}u_{{6}}u_{{8}}-a_{{18}}u_{{5}}u_{{8}}
   +a_{{19}}u_{{6}}u_{{7}}-a_{{20}}u_{{3}}u_{{8}}
   +a_{{20}}u_{{5}}u_{{7}}+a_{{21}}u_{{3}}u_{{7}}+\\
 &+a_{{14}}u_{{8}}
   -a_{{15}}u_{{7}}+a_{{17}}u_{{5}}+a_{{18}}u_{{4}}-a_{{19}}u_{{3}}
   -a_{{21}}u_{{1}}-a_{{11}}+2\,a_{{12}} \Big) u_{{9}}+\\
 &
 +a_{{17}}u_{{6}}u_{{7}}+a_{{18}}u_{{5}}u_{{7}}
 +a_{{20}}u_{{3}}u_{{7}}-a_{{14}}u_{{7}}
 -a_{{17}}u_{{3}}-a_{{18}}u_{{2}}-a_{{20}}u_{{1}}
 +a_{{9}}.
\end{align*}
The remaining 12 equations which we do not present here can be again solved by
consecutive integrations.

\subsection{$A_4$ algebra (Lie algebra of $SL(5,\C)$ group)}
In this case the decomposition (\ref{commdecomp}) of $\g$ has the following form
\begin{align*}
\ap_1\oplus&\ap_2\oplus\ap_3\oplus\ap_4\oplus
\h\oplus\am_4\oplus\am_3\oplus\am_2\oplus\am_1=\\
=&\spn\left\{X_1,\ldots,X_4\right\}
 \oplus\spn\left\{X_5,X_{6},X_{7}\right\}
 \oplus\spn\left\{X_{8},X_{9}\right\}
 \oplus\spn\left\{X_{10}\right\}\oplus\\
 &\oplus\spn\left\{X_{11},\ldots,X_{14}\right\}\oplus\\
 &\oplus\spn\left\{X_{15}\right\}
 \oplus\spn\left\{X_{16},X_{17}\right\}
 \oplus\spn\left\{X_{18},X_{19},X_{20}\right\}
 \oplus\spn\left\{X_{21},\ldots,X_{24}\right\},
\end{align*}
where the generators $X_i$ form an ordered basis fulfilling the
Definition~\ref{ordbdef}. The explicit parametrization of the algebra by these
generators reads:
\begin{displaymath}
\hspace{-2cm}
\sum_{i=1}^{24}a_iX_i=
\left[ \begin {array}{ccccc}
a_{{11}}&a_{{4}}&a_{{3}}&a_{{2}}&a_{{1}}\\a_{{21}}&-a_{{11}}+a_{{12}}&a_{{7}}&a_{{6}}&a_{{5}
}\\a_{{22}}&a_{{18}}&-a_{{12}}+a_{{13}}&a_{{9}}&a_{{
8}}\\a_{{23}}&a_{{19}}&a_{{16}}&-a_{{13}}+a_{{14}}&a
_{{10}}\\a_{{24}}&a_{{20}}&a_{{17}}&a_{{15}}&-a_{{14
}}\end {array} \right].
\end{displaymath}

\subsection{$B_4$ algebra (Lie algebra of $O(9,\C)$ group)}

In this case the decomposition (\ref{commdecomp}) of $\g$ has the following form
\begin{align*}
\ap_1\oplus&\ap_2\oplus\ap_3\oplus\ap_4\oplus
\h\oplus\am_4\oplus\am_3\oplus\am_2\oplus\am_1=\\
 =&\spn\left\{X_1,\ldots,X_7\right\}
 \oplus\spn\left\{X_{8},\ldots,X_{12}\right\}
 \oplus\spn\left\{X_{13},X_{14},X_{15}\right\}
 \oplus\spn\left\{X_{16}\right\}\oplus&\\
 &\oplus\spn\left\{X_{17},\ldots,X_{21}\right\}\oplus\\
 &\oplus\spn\left\{X_{22}\right\}
 \oplus\spn\left\{X_{23},X_{24},X_{25}\right\}
 \oplus\spn\left\{X_{26},\ldots,X_{30}\right\}
 \oplus\spn\left\{X_{31},\ldots,X_{36}\right\},
\end{align*}
where the generators $X_i$ form an ordered basis fulfilling the
Definition~\ref{ordbdef}. The explicit parametrization of the algebra by these
generators $\sum_{i=1}^{36}a_iX_i$ reads:
 { \small
\begin{displaymath}
\left[ \begin {array}{ccccccccc}
0&-a_{{33}}&-a_{{10}}&-a_{{14}}&-a_{{16}}&a_{{4}}&a_{{27}}&a_{{23}}&a_{{21}}\\-a_{{4}}&a
_{{17}}&a_{{1}}&a_{{2}}&a_{{3}}&0&a_{{7}}&a_{{6}}&a_{{5}}
\\-a_{{27}}&a_{{36}}&a_{{17}}-a_{{18}}&a_{{25}}&a_{{
26}}&-a_{{7}}&0&a_{{29}}&a_{{28}}\\-a_{{23}}&a_{{35}
}&a_{{12}}&a_{{18}}-a_{{19}}&a_{{22}}&-a_{{6}}&-a_{{29}}&0&a_{{24}}
\\-a_{{21}}&a_{{34}}&a_{{11}}&a_{{15}}&a_{{19}}-2\,a
_{{20}}&-a_{{5}}&-a_{{28}}&-a_{{24}}&0\\a_{{33}}&0&-
a_{{30}}&-a_{{31}}&-a_{{32}}&-a_{{17}}&-a_{{36}}&-a_{{35}}&-a_{{34}}
\\a_{{10}}&a_{{30}}&0&-a_{{8}}&-a_{{9}}&-a_{{1}}&-a_
{{17}}+a_{{18}}&-a_{{12}}&-a_{{11}}\\a_{{14}}&a_{{31
}}&a_{{8}}&0&-a_{{13}}&-a_{{2}}&-a_{{25}}&-a_{{18}}+a_{{19}}&-a_{{15}}
\\a_{{16}}&a_{{32}}&a_{{9}}&a_{{13}}&0&-a_{{3}}&-a_{
{26}}&-a_{{22}}&-a_{{19}}+2\,a_{{20}}\end {array} \right].
\end{displaymath}
}

\subsection{$C_4$ (Lie algebra of $Sp(8,\C)$ group)}

In this case the decomposition (\ref{commdecomp}) of $\g$ has the following form
\begin{align*}
\ap_1\oplus&\ap_2\oplus\ap_3\oplus\ap_4\oplus
\h\oplus\am_4\oplus\am_3\oplus\am_2\oplus\am_1=\\
 =&\spn\left\{X_1,\ldots,X_{10}\right\}
 \oplus\spn\left\{X_{11},X_{12},X_{13}\right\}
 \oplus\spn\left\{X_{14},X_{15}\right\}
 \oplus\spn\left\{X_{16}\right\}\oplus&\\
 &\oplus\spn\left\{X_{17},\ldots,X_{21}\right\}\oplus&\\
 &\oplus\spn\left\{X_{22}\right\}
 \oplus\spn\left\{X_{23},X_{24}\right\}
 \oplus\spn\left\{X_{25},X_{26},X_{27}\right\}
 \oplus\spn\left\{X_{28},\ldots,X_{36}\right\},
\end{align*}
where the generators $X_i$ form an ordered basis fulfilling the
Definition~\ref{ordbdef}. The explicit parametrization of the algebra by these
generators $\sum_{i=1}^{36}a_iX_i$ reads:
 { \small
\begin{displaymath}
\left[ \begin {array}{cccccccc}
a_{{17}}-a_{{18}}&a_{{24}}&a_{{25}}&a_{{26}}&a_{{10}}&a_{{9}}&a_{{7}}&a_{{6}}\\a_{{13}}&-
a_{{19}}+a_{{18}}&a_{{22}}&a_{{23}}&a_{{9}}&a_{{8}}&a_{{5}}&a_{{3}}
\\a_{{12}}&a_{{15}}&a_{{19}}-a_{{20}}&a_{{21}}&a_{{7
}}&a_{{5}}&a_{{4}}&a_{{2}}\\a_{{11}}&a_{{14}}&a_{{16
}}&a_{{20}}&a_{{6}}&a_{{3}}&a_{{2}}&a_{{1}}\\a_{{27}
}&a_{{28}}&a_{{30}}&a_{{31}}&-a_{{17}}+a_{{18}}&-a_{{13}}&-a_{{12}}&-a
_{{11}}\\a_{{28}}&a_{{29}}&a_{{32}}&a_{{34}}&-a_{{24
}}&a_{{19}}-a_{{18}}&-a_{{15}}&-a_{{14}}\\a_{{30}}&a
_{{32}}&a_{{33}}&a_{{35}}&-a_{{25}}&-a_{{22}}&-a_{{19}}+a_{{20}}&-a_{{
16}}\\a_{{31}}&a_{{34}}&a_{{35}}&a_{{36}}&-a_{{26}}&
-a_{{23}}&-a_{{21}}&-a_{{20}}\end {array} \right].
\end{displaymath}
}

\subsection{$D_4$ algebra (Lie algebra of $O(8,\C)$ group)}

In this case the decomposition (\ref{commdecomp}) of $\g$ has the following form
\begin{align*}
\ap_1\oplus&\ap_2\oplus\ap_3\oplus\ap_4\oplus
\h\oplus\am_4\oplus\am_3\oplus\am_2\oplus\am_1=\\
 =&\spn\left\{X_1,\ldots,X_6\right\}
 \oplus\spn\left\{X_7,\ldots,X_{10}\right\}
 \oplus\spn\left\{X_{11}\right\}
 \oplus\spn\left\{X_{12}\right\}\oplus\\
 &\oplus\spn\left\{X_{13},\ldots,X_{16}\right\}\oplus\\
 &\oplus\spn\left\{X_{17}\right\}
 \oplus\spn\left\{X_{18}\right\}
 \oplus\spn\left\{X_{19},\ldots,X_{22}\right\}
 \oplus\spn\left\{X_{23},\ldots,X_{28}\right\},
\end{align*}
where the generators $X_i$ form an ordered basis fulfilling the
Definition~\ref{ordbdef}. The explicit parametrization of the algebra by these
generators $\sum_{i=1}^{28}a_iX_i$ reads: { \small
\begin{displaymath}
\left[ \begin {array}{cccccccc}
a_{{13}}-a_{{16}}&a_{{17}}&a_{{21}}&a_{{22}}&0&a_{{6}}&a_{{5}}&a_{{4}}\\a_{{12}}&a_{{13}}
-a_{{14}}+a_{{16}}&a_{{19}}&a_{{20}}&-a_{{6}}&0&a_{{3}}&a_{{2}}
\\a_{{8}}&a_{{10}}&a_{{14}}-a_{{15}}&a_{{18}}&-a_{{5
}}&-a_{{3}}&0&a_{{1}}\\a_{{7}}&a_{{9}}&a_{{11}}&a_{{
15}}&-a_{{4}}&-a_{{2}}&-a_{{1}}&0\\0&-a_{{23}}&-a_{{
24}}&-a_{{25}}&-a_{{13}}+a_{{16}}&-a_{{12}}&-a_{{8}}&-a_{{7}}
\\a_{{23}}&0&-a_{{26}}&-a_{{27}}&-a_{{17}}&-a_{{13}}
+a_{{14}}-a_{{16}}&-a_{{10}}&-a_{{9}}\\a_{{24}}&a_{{
26}}&0&-a_{{28}}&-a_{{21}}&-a_{{19}}&-a_{{14}}+a_{{15}}&-a_{{11}}
\\a_{{25}}&a_{{27}}&a_{{28}}&0&-a_{{22}}&-a_{{20}}&-
a_{{18}}&-a_{{15}}\end {array} \right].
\end{displaymath}
}

\section{Implementation of the algorithm}

We have presented fully applicable algorithm for reducing the highly non linear
system of equations (\ref{eq-wn0}) for the parameters $u_k$ to a hierarchy of
Riccati matrix equations and integrals. The results presented in
Section~\ref{secresults} were obtained by computer program implementing this
algorithm written under \emph{Maple}. The existence of working computer program
proves the usability of the algorithm. The computation complexity of this
algorithm grows fast with the rank of the algebra $N$. The crucial point is
computation or the inverse (\ref{eq-wn3}), which has to be realized by Cramer's
rule of a computational complexity scaling as $d!$ with the dimension of the
inverted matrix $d$. Since the problem splits into subsystems corresponding to
decomposition (\ref{commdecomp}) the computation of the inverse (\ref{eq-wn3})
is substantially simplified, since it can be calculated by inverting the
matrices corresponding to each $\ap_k$ separately. Thus, the crucial ingredient
for the computation complexity turns out to be the inversion of the matrix
corresponding to $\ap_1$ which has the largest dimension. The dimension of
$\ap_1$ depends on the algebra type in the following way:
\begin{equation}
\renewcommand{\arraystretch}{1.3}
\begin{array}{r|c|c|c|c}
& A_N & B_N & C_N & D_N\\
\hline
\dim\ap_1 & N & 2N-1 & \frac12{N(N+1)} & 2N-2
\end{array}
\end{equation}
It is worth stressing that the behavior of $\dim\ap_1$ for $C_N$ differs from
the others since in grows quadratically with $N$ while for $A_N$, $B_N$ and
$D_N$ there is a linear growth of  $\dim\ap_1$ with $N$, so the computational
complexity grows much faster with $N$ for algebras $C_N$ than for the other
ones. On a standard PC the algorithm is executed in a reasonable time for
$\dim\ap_1\leqslant10$ and the last algebras in each series fulfilling this
condition are $A_{10}$, $B_5$, $C_4$ and $D_6$.

\section{The algebra $G_2$}

In the previous sections we restricted our considerations to the classical
simple Lie algebras. The choice was not dictated merely by the fact that
exceptional Lie algebras seem to play a less prominent role in various
applications. We show below some obstacles one encounters when trying to apply
the presented method to the $G_2$ algebra. We discus how the fact that the
$G_2$ algebra does not fulfill some of the lemmas proved for classical groups
affects the usability of Wei-Norman method in this case and to what extend some
of the results presented in this paper apply to $G_2$.

The lowest dimensional matrix representation of $G_2$ is 7-dimensional
\cite{humphreys}. We use the following explicit matrix representation of $G_2$:
\begin{equation}\label{g2par}
\sum_{i=1}^{14}a_iX_i=
\left[ \begin {array}{ccccccc} 0&-a_{{10}}\sqrt {2}&a_{{6}}\sqrt {2}&
a_{{3}}\sqrt {2}&-a_{{5}}\sqrt {2}&a_{{9}}\sqrt {2}&-a_{{12}}\sqrt {2}
\\a_{{5}}\sqrt {2}&a_{{7}}&a_{{4}}&a_{{1}}&0&-a_{{3}
}&a_{{6}}\\-a_{{9}}\sqrt {2}&-a_{{11}}&-a_{{7}}+a_{{
8}}&a_{{2}}&a_{{3}}&0&a_{{10}}\\a_{{12}}\sqrt {2}&a_
{{14}}&a_{{13}}&-a_{{8}}&-a_{{6}}&-a_{{10}}&0\\a_{{
10}}\sqrt {2}&0&a_{{12}}&a_{{9}}&-a_{{7}}&a_{{11}}&-a_{{14}}
\\-a_{{6}}\sqrt {2}&-a_{{12}}&0&a_{{5}}&-a_{{4}}&a_{
{7}}-a_{{8}}&-a_{{13}}\\-a_{{3}}\sqrt {2}&-a_{{9}}&-
a_{{5}}&0&-a_{{1}}&-a_{{2}}&a_{{8}}\end {array} \right],
\end{equation}
where $X_i$ are the root vectors, $\np=\spn\left\{X_1,\ldots,X_6\right\}$,
$\h=\spn\left\{X_7,X_8\right\}$, $\nm=\spn\left\{X_9,\ldots,X_{14}\right\}$ and
the root vectors corresponding to positive simple roots are $X_2$ (long) and
$X_6$ (short).

First obstacle was already mentioned in the proof of Lemma \ref{nilpot}. The
$G_2$ root system contains root strings or length 4, so the nilpotency order of
adjoint operators $\ad{X}$ may be equal to 4 for some root vectors $X$ and
$\exp(u\ad{X})$ for such vectors is a polynomial of order 3 in $u_i$. For
example in the parametrization (\ref{g2par}): {\small
\begin{displaymath}
\exp(u\ad{X_3})=
\left[ \begin {array}{cccccccccccccc} 1&0&0&0&-3\,u&-3\,{u}^{2}&0&0&0
&0&0&0&-{u}^{3}&0\\0&1&0&0&0&0&0&0&3\,u&-3\,{u}^{2}&0
&0&0&{u}^{3}\\0&0&1&0&0&0&0&-u&0&0&0&-{u}^{2}&0&0
\\0&0&0&1&0&0&0&0&0&0&0&0&0&0\\0&0
&0&0&1&2\,u&0&0&0&0&0&0&{u}^{2}&0\\0&0&0&0&0&1&0&0&0
&0&0&0&u&0\\0&0&0&0&0&0&1&0&0&0&0&u&0&0
\\0&0&0&0&0&0&0&1&0&0&0&2\,u&0&0\\0
&0&0&0&0&0&0&0&1&-2\,u&0&0&0&{u}^{2}\\0&0&0&0&0&0&0&0
&0&1&0&0&0&-u\\0&0&0&0&0&0&0&0&0&0&1&0&0&0
\\0&0&0&0&0&0&0&0&0&0&0&1&0&0\\0&0
&0&0&0&0&0&0&0&0&0&0&1&0\\0&0&0&0&0&0&0&0&0&0&0&0&0&
1\end {array} \right].
\end{displaymath}
} As a consequence, the equations obtained are no longer Riccati equations.

The second obstacle is that for $G_2$ group the decomposition
(\ref{commdecomp}) having the properties described by Lemma \ref{comma} and
Corollary \ref{cordecomp} does not exist. The rank of $G_2$ is 2, so one would
expect the decomposition of $\np$ into two commuting subalgebras generated by
some disjoint subsets of root vectors, but it turns out that this is not the
case. Since there are finitely many potential decompositions of this type it is
easy to check all of them. It turns out that the only nontrivial decomposition
of $\np$ into two subalgebras is
\begin{equation}
\np=\ca\oplus\cb:=\spn\left\{X_1,X_2,X_3\right\}\oplus\spn\left\{X_4,X_5,X_6\right\},
\end{equation}
where $X_i$ are defined in (\ref{g2par}). It easy to check that $\ca$ is
commutative, but $\cb$ is not. Moreover $\ca$ is not an ideal in $\np$. Thus,
both algebras $\ca$ and $\cb$ do not fulfill the Lemma \ref{comma}. As a
consequence the invariance properties described by Lemma \ref{invsp} are
lacking and the system (\ref{eq-wn2a}) will not separate into blocks smaller
than those corresponding to the decomposition $\g=\nm\oplus\h\oplus\np$. This
is the only decomposition that survives.

The equations corresponding to sector $\np$ are:
\begin{align*}
u^{\prime}_{{1}}=&
 2\,a_{{14}}u_{{2}}{u_{{5}}}^{3}
  +3\,a_{{14}}{u_{{3}}}^{2}{u_{{5}}}^{2}
  +6\,a_{{10}}u_{{3}}{u_{{5}}}^{2}+2\,a_{{11}}{u_{{5}}}^{3}
  +3\,a_{{12}}u_{{2}}{u_{{5}}}^{2}
  +a_{{13}}{u_{{3}}}^{3}
  +\\  &
  -3\,a_{{6}}{u_{{3}}}^{2}+3\,a_{{9}}{u_{{5}}}^{2}
  -3\,a_{{12}}u_{{1}}u_{{3}}-a_{{13}}u_{{1}}u_{{2}}
  -a_{{14}}{u_{{1}}}^{2}
  +a_{{4}}u_{{2}}+3\,a_{{5}}u_{{3}}
  +\\  &
  + \left( a_{{7}}+a_{{8}} \right) u_{{1}}
  +a_{{1}},
\\
u^{\prime}_{{2}}=&
 -a_{{14}}{u_{{3}}}^{3}-3\,a_{{10}}{u_{{3}}}^{2}
 -3\,a_{{12}}u_{{2}}u_{{3}}-a_{{13}}{u_{{2}}}^{2}
 -a_{{14}}u_{{1}}u_{{2}}-3\,a_{{9}}u_{{3}}
 -a_{{11}}u_{{1}}
  +\\  &
 + \left( -a_{{7}}+2\,a_{{8}} \right) u_{{2}}+a_{{2}},
\\
u^{\prime}_{{3}}=&
 a_{{14}}u_{{2}}{u_{{5}}}^{2}+2\,a_{{14}}{u_{{3}}}^{2}u_{{5}}
 +4\,a_{{10}}u_{{3}}u_{{5}}+a_{{11}}{u_{{5}}}^{2}
 +2\,a_{{12}}u_{{2}}u_{{5}}-a_{{12}}{u_{{3}}}^{2}
 -a_{{13}}u_{{2}}u_{{3}}
  +\\  &
 -a_{{14}}u_{{1}}u_{{3}}
 +a_{{6}}u_{{2}}+a_{{8}}u_{{3}}+2\,a_{{9}}u_{{5}}
 -a_{{10}}u_{{1}}+a_{{3}},
\\
u^{\prime}_{{4}}=&
 a_{{14}}u_{{2}}{u_{{4}}}^{2}+3\,a_{{14}}u_{{3}}u_{{4}}u_{{5}}
 -a_{{14}}{u_{{5}}}^{3}+3\,a_{{10}}u_{{4}}u_{{5}}
 +a_{{11}}{u_{{4}}}^{2}-3\,a_{{12}}{u_{{5}}}^{2}
 +a_{{13}}u_{{2}}u_{{4}}
 +\\  &
 +3\,a_{{13}}u_{{3}}u_{{5}}
 -a_{{14}}u_{{1}}u_{{4}}-3\,a_{{6}}u_{{5}}-a_{{13}}u_{{1}}
 + \left( 2\,a_{{7}}-a_{{8}} \right) u_{{4}}+a_{{4}},
\\
u^{\prime}_{{5}}=&
 a_{{14}}u_{{2}}u_{{4}}u_{{5}}+a_{{14}}{u_{{3}}}^{2}u_{{4}}
 +a_{{14}}u_{{3}}{u_{{5}}}^{2}+2\,a_{{10}}u_{{3}}u_{{4}}
 +a_{{10}}{u_{{5}}}^{2}+a_{{11}}u_{{4}}u_{{5}}
 +a_{{12}}u_{{2}}u_{{4}}
 +\\  &
 -a_{{12}}u_{{3}}u_{{5}}
 +a_{{13}}{u_{{3}}}^{2}-a_{{14}}u_{{1}}u_{{5}}
 -2\,a_{{6}}u_{{3}}+a_{{7}}u_{{5}}+a_{{9}}u_{{4}}
 -a_{{12}}u_{{1}}+a_{{5}},
\\
u^{\prime}_{{6}}=&
 a_{{14}}u_{{2}}u_{{5}}{u_{{6}}}^{2}
 +a_{{14}}{u_{{3}}}^{2}{u_{{6}}}^{2}
 +2\,a_{{10}}u_{{3}}{u_{{6}}}^{2}+a_{{11}}u_{{5}}{u_{{6}}}^{2}
 +a_{{12}}u_{{2}}{u_{{6}}}^{2}
 +a_{{14}}u_{{2}}u_{{4}}u_{{6}}
 +\\  &
 +a_{{14}}u_{{3}}u_{{5}}u_{{6}}+a_{{9}}{u_{{6}}}^{2}
 +a_{{10}}u_{{5}}u_{{6}}+a_{{11}}u_{{4}}u_{{6}}
 +a_{{12}}u_{{3}}u_{{6}}
 +a_{{13}}u_{{2}}u_{{6}}
 -a_{{14}}u_{{3}}u_{{4}}
 +\\  &
 +a_{{14}}{u_{{5}}}^{2}
 -a_{{10}}u_{{4}}+2\,a_{{12}}u_{{5}}-a_{{13}}u_{{3}}
 + \left( a_{{7}}-a_{{8}} \right) u_{{6}}+a_{{6}},
\end{align*}
The polynomials on the right hand side of above system are of order 4 and not
only 3. The maximal power of single variable is equal to 3, but the lack of
decomposition into commuting subalgebras causes the appearance of higher order
terms, because the product of two noncommuting operators of nilpotency order 4
does not have to be of the same nilpotency order.

The remaining eight functions $u_i$ corresponding to sectors $\h$ and $\nm$ can
be computed by consecutive integration, provided the solutions of the above six
equations are found. This property of the decomposition
$\g=\nm\oplus\h\oplus\np$ holds also in this case.

The question of applicability of our method to other exceptional Lie groups is
also interesting and definitely worth answering. The root systems of other
exceptional Lie groups do not contain root strings of order 4, but the problem
of existence of the decompositions into sum of commuting subalgebras with
relevant invariance properties deserves deeper study and will be provided
elsewhere.

\section{Discussion and remarks}

The general method for solving the matrix Riccati equation is not known, so the
method presented in this paper does not provide the general solution of  the
system (\ref{precession}-\ref{Mdec}). Nevertheless there are many methods to
study matrix Riccati equations (see for example \cite{carinena12} and
references therein), so our method yields a major reduction of complexity of
the original problem (\ref{precession}-\ref{Mdec}).

Scalar and matrix Riccati equations and system of the form
(\ref{precession}-\ref{Mdec}) are examples of the so called \emph{Lie systems}.
The theory of such systems is still being developed, see for example
\cite{winternitz83} and the most recent review paper \cite{carinena12} with
exhaustive list of references therein. It is known that every system of Riccati
equations is related to a Lie group action and solution of this system is
equivalent to solution of the system of the form (\ref{precession}-\ref{Mdec}),
but not every system of the form (\ref{precession}-\ref{Mdec}) is equivalent to
Riccati equation system. In \cite{charzynski13} we have shown that for
$SU(N+1)$ and $SL(N+1,\C)$ the system (\ref{precession}-\ref{Mdec}) is
equivalent to hierarchy of Riccati equations and linear equations. Here we have
shown that the same statements holds for all classical Lie groups, but can not
be generalized to all Lie groups, presenting $G_2$ as a counterexample.

\section*{Acknowledgments}

We gratefully acknowledge fruitful discussions with Javier de Lucas. The
presented results are obtained in frames of the the Polish National Science
Center project MAESTRO DEC-2011/02/A/ ST1/00208 support of which is gratefully
acknowledged by both authors.

\end{document}